# Strategies and Performances of Soft Input Decryption

**Natasa Zivic**[†]
*natasa.zivic@uni-siegen.de*
University of Siegen, Hoelderlinstrasse 3, Siegen, Germany

**Summary**

This paper analyzes performance aspects of Soft Input Decryption and *L*-values. Soft Input Decryption is a novel method which uses *L*-values (soft output) of a SISO channel decoder for the correction of input of Soft Input Decryption (SID blocks) which have been modified during the transmission over a noisy channel. The method is based on the combination of cryptography and channel coding improving characteristics of both of them. The algorithm, strategies and scenarios of Soft Input Decryption are described.

The results of the tested performance of *L*-values show how many *L*-values are necessary for the correction of SID blocks. This number is higher than the number of wrong bits. This difference is presented. The number of *L*-values is estimated for different lengths of SID blocks as well as for different $E_b/N_0$ ratios. Space characteristics of *L*-values are not analyzed, because they depend on the used SISO decoding algorithm. In this paper, Maximum A-Posteriori (MAP) algorithm is used.

The time performance of Soft Input Decryption depends on the used cryptographic mechanism for the verification of the cryptographic check values (digital signatures (based on Elliptic Curve Cryptography), MACs (based on CBC-DES) and H-MACs (based on SHA-1)).

The combination of the performance of *L*-values and time performance gives an estimation, if and when Soft Input Decryption can be performed in practice. Further optimizations are outlined.

## 1. Introduction

The combination of a SISO decoder and Decryptor, which is analyzed in this paper, can be considered as a concatenation of codes: en-/decryptor has the role of an outer en-/decoder and the channel encoder/SISO channel decoder has the role of an inner en-/decoder has (Fig. 1).

Concatenation of codes, presented as an outer and inner code was already devised by Forney in 1966 [1]. In literature, it is known as concatenated codes [2], general concatenated codes [3] or super channel codes. The inner code is generally short and decoded with a soft decision decoding algorithm, while the outer code is generally longer and decoded with an algebraic decoding method [2].

In most cases a convolutional code is used as an inner code in combination with a Reed Solomon code or another convolutional code as an outer code. Such a type of concatenated codes can be compared to the combination of codes investigated in this work (Fig.1). Two good characteristics are the result of such a concatenated schema: good error performance because of the use of SISO principle and good security performance as the result of the use of the cryptographic mechanisms.

The next common point of this work with previous works in coding is the idea of the use of reliability in decoding. There are several works which explore the reliability-based soft-decision decoding algorithms for linear block codes, using the concept of error correction by ordering the decoded bits by their reliability values. The values of soft outputs of the decoder have been used as reliability values.

The idea of inversion of the least probable bits (with the lowest reliability values) originated from Chase decoding algorithms [4] in 1972, which were the generalization of the GMD (Generalized Minimum Distance) algorithms from 1966 [1]. These algorithms have been applied to a binary (n, k) linear block code and are referenced as LRP (Least Reliability Positions) algorithms.

Chase algorithms generate a list of candidate code words by complementing all possible combinations of bits with the lowest reliability values. The candidate with the best metric is the decoded solution.

The similarity to the method of the Soft Input Decryption, is the use of *L*-values reordered and iteratively tested. The difference is that Soft Input Decryption uses two decoders (inner and outer) and a non-linear block code (cryptographic algorithms). Two codes enable the use of feedback from the outer to inner code.

The next group of algorithms, which use decoding based on ordering of L-values, is a group of MRIP (Most Reliable Independent Positions) - reprocessing decoding algorithms, as the Most Reliable Basis [MRB], the Least Reliable Basis [LRB] and the Ordered Statistic Decoding Algorithm [2].

Joint source channel coding is the another topic related to this work. The cooperation between the source and channel decoder enables a better use of information of both decoders and better decoding results [5]. It is based on the turbo – principle, as well as Softbit - Source Decoding [6] and Iterative Source – Channel decoding [7]. The similarity to Soft Input Decryption is the use of iterative information exchange between the two elements of the receiver: channel and source decoder, in case of joint source channel coding, rsp. channel decoder and decryptor in case of Soft Input Decryption.

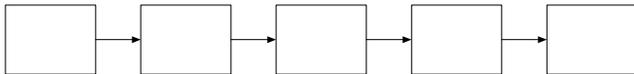

Fig. 1 Communication System using Concatenated Codes.

## 2. Soft Input Decryption Algorithm

Soft Input Decryption [8] is a method for the correction of SID blocks which contain cryptographic check values (digital signatures, MACs, H-MACs) by using *L*-values as the output of the SISO channel decoder. Cryptographic check values provide data integrity, data origin authentication and non repudiation [9].

Soft Input Decryption works block oriented. The input for Soft Input Decryption contains data which are secured by cryptographic check values. The block which has to be corrected by Soft Input Decryption after channel decoding is called SID block (Soft Input Decryption block). It may contain data and cryptographic check values, or just cryptographic check values, depending on the used scenario (see Section 2.4).

In Fig. 2 the standard verification process without Soft Input Decryption is presented.

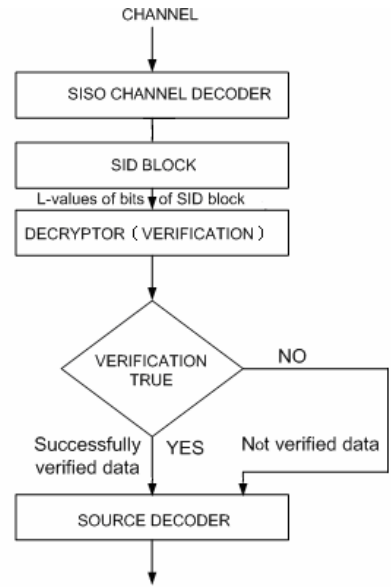

Fig. 2 Verification of a SID block without Soft Input Decryption.

The algorithm of Soft Input Decryption (Fig. 3) is as follows:

The Soft Input Decryption is successfully completed, if the verification of the cryptographic check value is successful, i.e. the output is "true". If the verification is negative, the soft output of the channel decoder is analyzed and the bits with the lowest |*L*|-values are flipped (XOR 1). Then the decryptor performs the verification process and proves the result of the verification again. If the verification is negative, bits with another combination of the lowest |*L*|-values are changed. This iterative process will stop when the verification is successful or the needed resources are consumed.

In the case that the attempts for correction fail, the number of modified bits is too large as a result of a very noisy channel, a very long SID block or an attack, so that the resources are not sufficient to find the correct content of a SID block.

It may happen that the attempts for the correction of a SID block succeed, but the content of a SID block is not equal to the original one: a collision happened. This case has a negligible probability if the length of the cryptographic check values are chosen under security aspects (for example, considering the "birthday paradox").

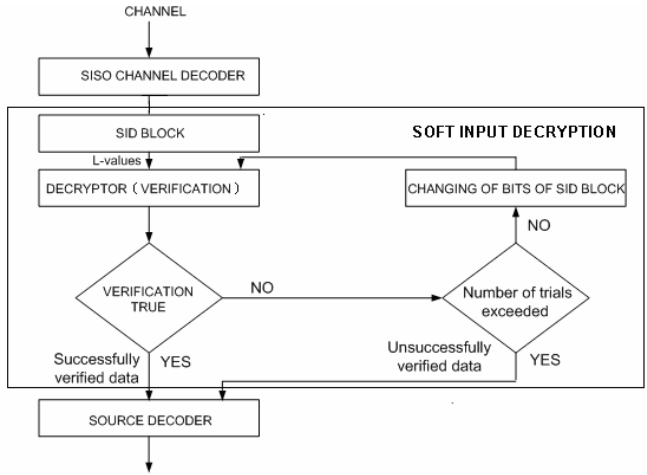

Fig. 3 Algorithm of Soft Input Decryption.

## 3. Simulations and Results

The chosen length of the SID block is 320 bit. The reason for this length of a SID block is that this is a size of a digital signature based on Elliptic Curve Cryptography over $GF(2^{160})$ or $GF(p)$ with $ld(p) = 160$. Further on BPSK modulation, the model of an AWGN channel, as well as a convolutional and a turbo code [10] are used in the simulations. The used convolutional encoder has a code rate $r = 1/2$ and a constraint length $m = 2$. The turbo encoder with $r = 1/3$ is based on two parallel RSC convolutional encoders with $r = ½$ and a block interleaver of depth 17 [12]. The turbo decoder performs 3 iterations. The decoder (convolutional and turbo) uses a MAP algorithm [11]. The results in Fig. 4 are shown for $E_b/N_0 > 1.5$ dB. The simulations have been programmed in C/C++ programming language. For each point of the curves 50.000 tests are performed which are sufficient to be representative.

*CCER* (Cryptographic Check Error Rate) is defined as:

$$CCER = \frac{number\ of\ incorrect\ SID\ blocks}{number\ of\ received\ SID\ blocks} \quad (1)$$

An incorrect SID block is a SID block which could not be successfully verified or a SID block, which has been falsely successfully verified, i.e. a collision happened. Vice versa, a correct SID block is a successfully verified SID block and no collision happened, i.e. it is identical to the sent SID block.

The complement of *CCER* is:

$$\overline{CCER} = 1 - CCER = \frac{number\ of\ correct\ SID\ blocks}{number\ of\ received\ SID\ blocks} \quad (2)$$

*CCER* is presented in Fig. 4 in relation to $E_b/N_0$. Soft Input Decryption is performed for trials up to the 8, rsp. the 16 lowest $|L|$-values (i.e. $2^8$ rsp. $2^{16}$ trials).

The Soft Input Decryption gain [dB] means the reduction of *CCER* by Soft Input Decryption depending on $E_b/N_0$. SID blocks can be successfully transmitted over a communication channel with $E_b/N_0$ of 2.5 dB using Soft Input Decryption. For example, if 1 of 10 SID blocks cannot be verified (at 4 dB) without Soft Input Decryption, all SID blocks can be corrected in the case of convolutional codes. In case of a turbo code *CCER* can be reduced from 1/30 to 0 (at 2.5 dB).

The results in Fig.5 are achieved for different lengths of SID blocks using convolutional coding: 128, 160, 320 and 384 bits, with up to 16 flipped bits corresponding to the lowest positions of $|L|$-values. As expected, the coding gain of Soft Input Decryption is influenced by the length of a SID block: it is lower when a SID block is longer.

In the following figures and sections a convolutional code is used only, because the simulations of convolutional codes are faster than of turbo codes and the coding gains of turbo codes are similar to those of convolutional codes (compare to Fig. 4).

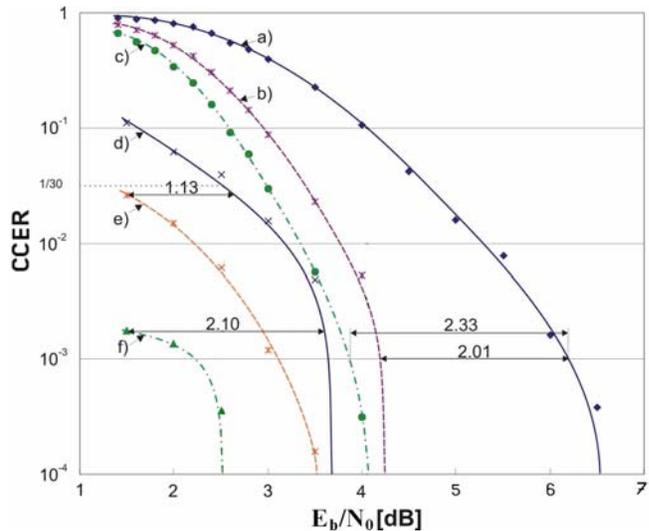

Fig. 4 Decrypting gain of 320 bit SID block [8]

a) Convolutional code without Soft Input Decryption
b) Convolutional code with Soft Input Decryption using up to the 8 lowest $|L|$-values
c) as b.), but using up to the 16 lowest $|L|$-values

d) Turbo code without Soft Input Decryption
e) Turbo code with Soft Input Decryption using up to the 8 lowest /L/-values
f) as e.), but using up to the 16 lowest /L/-values

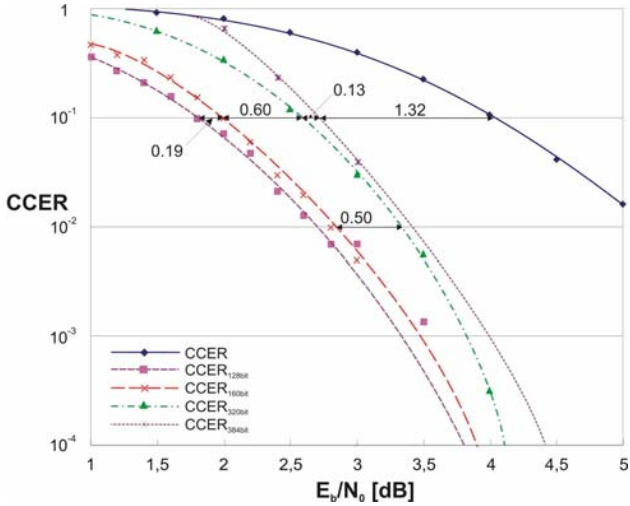

Fig. 5 *CCER* Coding gain of Soft Input Decryption using up to 16 lowest |L|-values, for different lengths of SID blocks

## 4. Strategies of Soft Input Decryption

The module "Changing of bits of SID block" (see Fig. 3) contains the strategy, in which sequence bits and combinations of bits of the SID block are changed, before the next verification is achieved. Depending on the strategy of Soft Input Decryption, different schedules of bit correction are possible.

The static strategy of Soft Input Decryption is used for the results of Chapter 3. Proposals for other strategies are also given in this Chapter.

### 4.1 Static Strategy

If the first verification after starting Soft Input Decryption is not successful, the bit with the lowest /L/-value of the SID block is flipped, assuming that the wrong bits are probably those with the lowest /L/-values. If the verification is again not successful, the bit with the second lowest /L/-value is changed. The next try will flip the bits with the lowest and second lowest /L/-value, then the bit with the third lowest /L/-value, etc. The process is limited by the number of bits with the lowest /L/-values, which should be tested. The strategy follows a representation of an increasing binary counter, whereby the lowest bit corresponds to the bit with the lowest /L/-value, etc.

The strategy is defined by following algorithm.

The static strategy orders the bits of the SID block by their /L/-values starting with the lowest one and monotonly increasing. The output of the sort algorithm is represented as an example in Fig. 6. $j$ is the increasing sequence of positions of /L/-values and $P_j$ indicates the position of the bit with the $j^{th}$ lowest /L/-value in the original SID block. The length of the SID block is $w$.

Fig. 6 Sorted sequence of bits of a SID block (an example).

The function "Changing of bits of SID block" of the Soft Input Decryption algorithm (Fig. 2) is shown in Fig. 7 in detail.

To control the strategy, an incrementing counter $i = 1, \ldots, 2^{N_{max}} - 1$ is used in binary representation of fixed lengths of $N_{max}$ with coefficients $c_{ij} \in \{0, 1\}$.

$$i = \sum_{j=1}^{N_{max}} c_{i,j} 2^{j-1} \quad (3)$$

$N_{max}$ is the maximum number of bits to be flipped, rsp. $2^{N_{max}} - 1$ is the maximum number of trials, if all verifications fail. Each value of the counter $i$ describes one trial. The indices $j$ of those coefficients $c_{ij}$ which are marked ($c_{ij} = 1$) indicate the positions of $L$-values of the bits to be flipped. These positions can be found by using the sorted table as in Fig. 6.

The sorted sequence $P_j$ can be limited to $P_{Nmax}$. $i$ is reset to 0 at the beginning of Soft input Decryption.

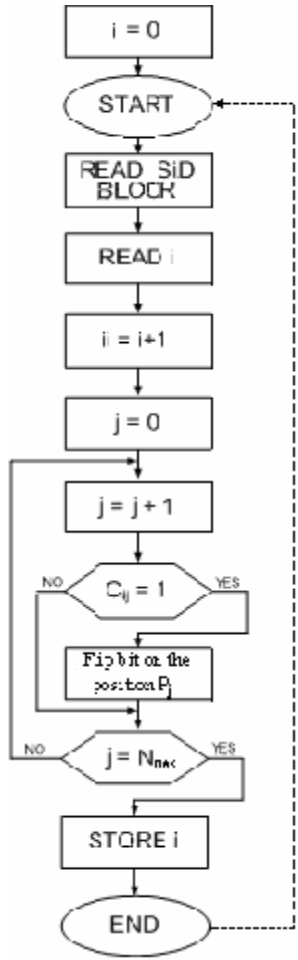

Fig. 7 Algorithm of the static strategy

The strategy is based on the following assumption: if bits are wrong decoded by the channel decoder, than they have the lowest /$L$/-values. Unfortunately, this assumption is not true, because $L$-values are probability based and give only an orientation which bits may be wrong decoded. It may happen, for example, that combinations of up to 7 bits have to be tested, when only 3 bits are wrong.

## 4.2 Dynamic Strategy

The static strategy sorts the $L$-values of single bits. The dynamic strategy calculates the $L$-values of groups of bits to decide which trial should be performed next. It can happen, that a group of bits result in a lower /$L$/-value than a /$L$/-value of a single bit, i.e. that a specific group of bits is probably wrong. By this way, it is possible to find the group of wrong bits in only a few trials, not testing all combinations of flipped bits untill the bits untill the right combination is found like in the static strategy. The calculation is based on use of the $L$ algebra [12] and much more complex than the static strategy. The elaboration of the dynamic strategy is for further study.

## 4.3 BER Based Strategy

The BER based strategy analyses, which number of errors is the most probable in the SID block, under consideration of $E_b/N_0$ rsp. BER. To reduce the potential number of tests, $L$-values are taken into account. Example: if $N_{max}$ is 16 and 4 bit errors are most probable for SID blocks of 320 bits, then up to $\binom{16}{4}$ instead of $\binom{320}{4}$ tests are performed. If $\binom{16}{4}$ tests are not successful, $\binom{16}{3}$ tests of 3 bit errors and $\binom{16}{5}$ tests of 5 bit errors have to be tested. The BER strategy is also for further study.

## 5. Application Scenarios

### 5.1 Scenario 1

In scenario 1 (Fig. 7) digital signatures giving message recovery are used [13]. In this type of digital signatures the digital signature is not appended to the message, but contains the message, practically the digital signature replaces the message.

If the verification process is successful, the message is also correctly recovered. The length of the message including redundancy is limited by the size of the underlying mathematical structure of the used asymmetric cryptography. For example, the length has to be shorter than 1024 bits, if RSA is used with a length of the modulus of 1024 bits, or shorter than 160 bits, if ECC (elliptic curve cryptography) is used over GF ($2^{160}$). Algorithms can be found in [13] and [14].

This scenario can be often found in transaction oriented applications exchanging short messages, which have to be authentic. Typical examples are measurement values in industrial metering systems (electricity, water, gas etc.), data generated by sensors, as well as stock exchange rates.

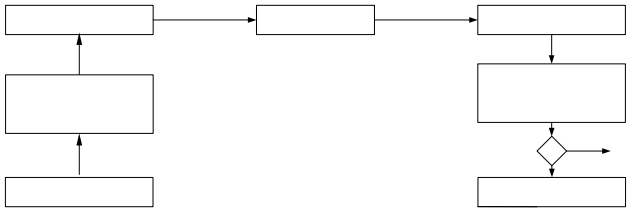

Fig. 7 Scenario of signatures giving message recovery

## 5.2 Scenario 2

Scenario 2 (Fig. 8) considers signatures with appendix [15]. Messages signed by signatures with appendix have arbitrary lengths, but Soft Input Decryption can only be applied to SID blocks of limited lengths (Note: the application of a collision free one-way hash function is considered here to be part of the signature generation and verification process). In this scenario, it is assumed that the message and the signature are transmitted separately and the SID block consists only of the signature. No message is transmitted via a communication channel different from the one used for the transmission of the digital signature (outband), or via the same communication channel (inband). The message itself is transmitted first and – if modified during transmission – corrected by redundancy within the message, by repetition or by agreement of the communication partners. So, it is assumed that finally the message is received correctly by the receiver. The digital signature is transmitted afterwards, either when it is generated on request or when the action described by the message should be executed.

Typical examples are legal contracts or bank transactions, which are prepared in advance, and the digital signature is transmitted at the requested moment. If an error occurs during the signature verification, the signature is not correct: it has been manipulated or errors occurred during the transmission, which can not be corrected by the channel decoder.

In this case, the Soft Input Decryption block consists of one signature block used for verification of the message.

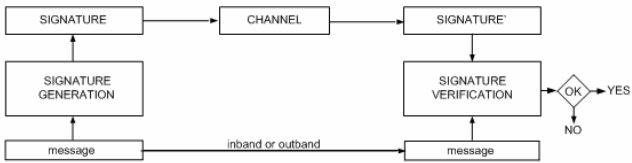

Fig. 8 Scenario of signatures with appendix

## 5.3 Scenario 3

Scenario 3 (Fig. 9) is the most general scenario and considers messages with a cryptographic check value: MAC [16], H-MAC [17] or digital signatures with appendix.

In this case, the SID block consists of a message extended by a cryptographic check value.

Typical examples are credit/debit applications and other bank applications.

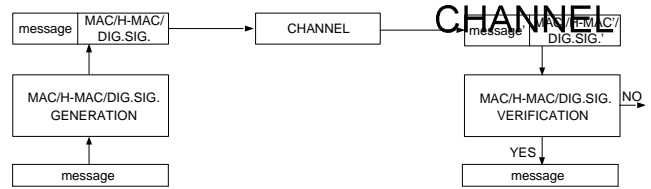

Fig. 9 Scenario of messages with cryptographic check values

At one hand, from the point of view of Soft Input Decryption it does not matter which scenario is applied. Soft Input Decryption always tries to correct the SID block as long as the verification fails. At other hand it is interesting to show that there are many aplications for Soft Input Decryption, even if the lengths of SID blocks are relativly short.

## 6. Performance aspects of L-values

6.1. The number of L-values required for correction

If the performance of Soft Input Decryption should be estimated, at first it is necessary to consider the reliability of the soft input.

Fig. 10 shows the number of *L*-values needed for correction of 95 % of SID blocks for different lengths of SID blocks: w = 128, 160, 320 and 384.
In case of SID blocks of length of 320 bits the usage of up to 8 *L*-values is sufficient for the correction if $E_b/N_0 > 4$ dB. The number of *L*-values necessary for correction increases exponentially with decreasing of S/N: for $E_b/N_0 < 3.5$ dB 13 *L*-values are needed for 95 % correction.

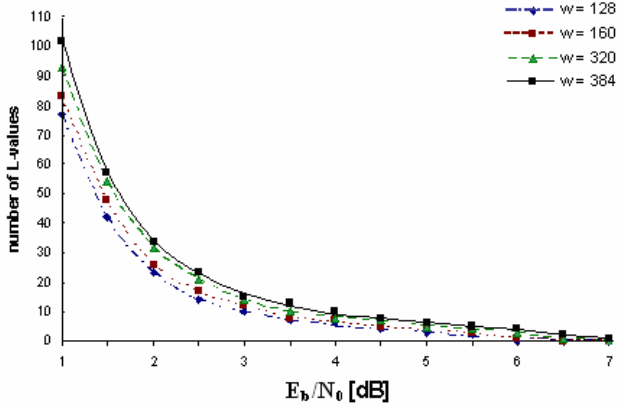

Fig. 10 Number of *L*-values needed for correction of 95 % of SID blocks

### 6.2. The theoretical minimum needed for correction

The problem is that reliability values themselves are not reliable. So, the question arises, which is the number of errors in a SID block, which is the lower limit of *L*-values to be used by Soft Input Decryption.

The probability $P_{w,i}$, that a code word of length $w$ – in this case a SID block – contains $i$ errors is:

$$P_{w,i} = \binom{w}{i} p^i (1-p)^{w-i} \quad (4)$$

where *p* represents BER.

Soft Input Decryption tests all possible combinations of up to $N_{max}$ *L*-values. The probability $P_w$ that, after correction by Soft Input Decryption, only errors which are longer than $N_{max}$ will not be corrected is [18]:

$$P_w = \sum_{i=N_{max}+1}^{w} \binom{w}{i} p^i (1-p)^{w-i} \quad (5)$$

Soft Input Decryption corrects the data received from channel decoding. Therefore *p* means the bit error rate after the channel decoder. *p* depends on the quality of a MAP decoder.

In the case of 95 % corrected SID blocks, $P_w = 0.05$.

$$0.05 \geq P_w = \sum_{i=N_{max}+1}^{w} \binom{w}{i} p^i (1-p)^{w-i} = 1 - \sum_{i=0}^{N_{max}} \binom{w}{i} p^i (1-p)^{w-i} \quad (6)$$

$N_{max}$ is calculated for the correction of more than 95 % errors has been done for lengths of SID blocks of $w = \{128, 160, 320, 384\}$ bits (Fig. 11).

$N_{max}$ in (5) and (6) means the minimum $N_{max}$ because the case considers the number of L-values for correction of $P_w$ wrong bits.

The results are shown in Fig. 11.

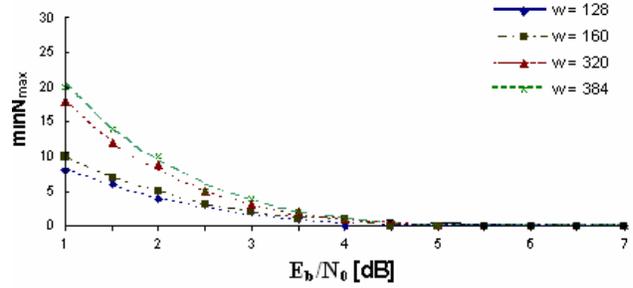

Fig. 11 Theoretically minimum number of bits ($minN_{max}$) to be changed

Now, the results of Fig. 10 and 11 can be compared. Fig. 12 shows the comparison for SID block of 320 bits. The reason for the significant difference is the incorrect allocation of /*L*/-values compared to wrong bits, due to imperfections in the implementation of MAP decoding algorithm.

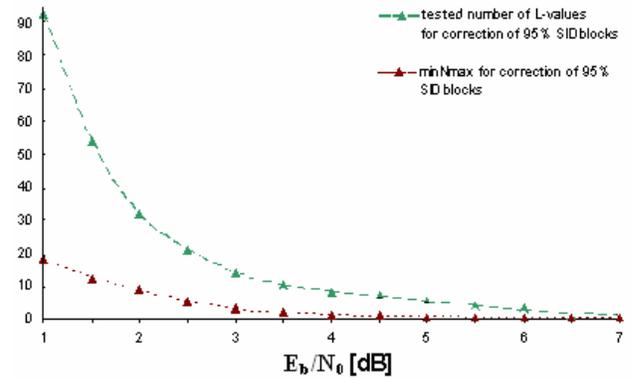

Fig. 12 Comparison of needed and theoretical number of *L*-values for SID block of the length 320 bits

### 6.3. Estimation of the number of L-values for the correction of long SID blocks

In this section the number of needed L-values for the correction of SID blocks longer than 384 bits is considered, because the application of Soft Input Decryption of longer SID blocks seems to be very interesting (see Scenario 3 of Section 2.4 or consider RSA digital signatures instead of ECC which will be discussed later in this section). The estimation for the number of required /L/-values is done by interpolation of the results for SID blocks of lengths of 128, 160 and 320 bits and extrapolation for longer SID blocks. The results for 384 bits known by simulations are used for the verification of the extrapolation.

Fig. 13 (w = 320, $E_b/N_0$ = 3 dB) shows the percentage of corrected SID blocks per /L/-value: 20 % of SID blocks are corrected with the lowest /L/-value. Additional 13 % are corrected with the second lowest /L/-value; with the third lowest /L/-value additional 11 % etc…; with the 15th lowest /L/-value additional 1 % more SID blocks are corrected than with the 14th lowest /L/-value.

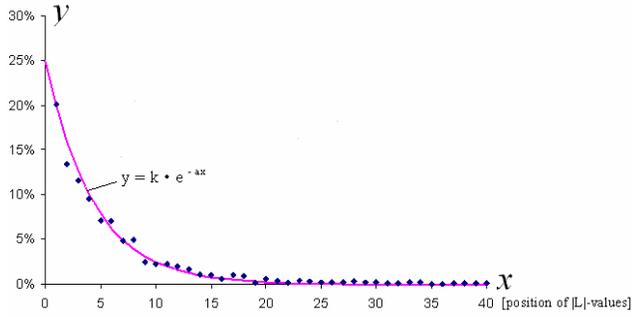
Fig. 13 The percentage of corrected SID blocks per /L/-value

The percentage of corrected SID blocks per /L/-value is tested for lengths of 128, 160 and 320 bits and for different $E_b/N_0$. The results of tests show curves with a negative exponential function of the form (see Fig. 13):

$$y = k\, e^{-ax} \quad (7)$$

Coefficients $k$ and $a$ depend on the length w of the SID block and $E_b/N_0$. This dependence is derived in this section and then applied for other values.

As the total percentage of corrected SID blocks is 100 % after testing of all L-values, the following condition is valid:

$$\sum_{i=1}^{w} y_i = k \sum_{i=1}^{w} e^{-ai} = \frac{ke^{-a}(e^{-aw}-1)}{e^{-a}-1} = 1 \quad (8)$$

or, by presenting $e^{-ai}$ as an infinite sum:

$$k = \frac{e^a - 1}{1 - e^{-aw}} = \frac{1 + a + \frac{a^2}{2!} + \frac{a^3}{3!} + \ldots - 1}{1 - e^{-aw}} = \frac{a + \frac{a^2}{2!} + \frac{a^3}{3!} + \ldots}{1 - e^{-aw}} = \frac{\sum_{i=1}^{\infty} \frac{a^i}{i!}}{1 - e^{-aw}} \approx \frac{a}{1 - e^{-aw}} \quad (9)$$

The negative exponential function is derived for w = {128, 160, 320} and $E_b/N_0$ = [1, 5.5] dB in steps 0.5 dB. For each function $a$ and $k$ are derived approximately and shown in Table 2.

The relation between the exponent $a$ and $E_b/N_0$ is observed in the next step. $S/N$ is used instead of $E_b/N_0$, respecting the code rate of ½:

$$\frac{S}{N} = \frac{E_b}{N_0} - 3dB \quad (10)$$

Approximations are done using $S/N$ ratio which is easier for calculation. The entries of the first and third row of Table 2 is used for approximation by a second order polynomial:

$$a = A\left(\frac{S}{N}\right)^2 + B\left(\frac{S}{N}\right) + C \quad (11)$$

The calculated coefficients A, B and C are shown in Table 2 for SID blocks of length of 128, 160 and 320 bits. The approximation of the dependence of $a$ on $S/N$ is shown in Fig. 14.

The coefficients A, B and C have different values dependent on the length of SID blocks. A, B and C show a linear relation to $w$. Therefore A, B and C are approximated by linear functions of $w$.

$$A = A(w) = KA \cdot w + NA \quad (12)$$

$$B = B(w) = KB \cdot w + NB \quad (13)$$

$$C = C(w) = KC \cdot w + NC \quad (14)$$

The values of the linear coefficients are shown in Table 3.

|  | $E_b/N_0$ [dB] | 1 | 1.5 | 2 | 2.5 | 3 | 3.5 | 4 | 4.5 | 5 |
|---|---|---|---|---|---|---|---|---|---|---|
| w = 128 | k | 0.083 | 0.127 | 0.271 | 0.323 | 0.418 | 0.461 | 0.646 | 0.733 | 0.821 |
|  | a | 0.08 | 0.12 | 0.24 | 0.28 | 0.35 | 0.38 | 0.5 | 0.55 | 0.6 |
| w = 160 | k | 0.030 | 0.078 | 0.127 | 0.221 | 0.284 | 0.47 | 0.582 | 0.792 | 1.059 |
|  | a | 0.038 | 0.075 | 0.12 | 0.2 | 0.25 | 0.38 | 0.46 | 0.6 | 0.7 |
| w = 320 | k | 0.02 | 0.025 | 0.062 | 0.221 | 0.258 | 0.419 | 0.733 | 1.222 | 1.454 |
|  | a | 0.02 | 0.025 | 0.06 | 0.2 | 0.23 | 0.35 | 0.5 | 0.8 | 0.9 |

Table 1 Coefficients *k* and *a* for SID block of *w* = 128, 160 and 320 bit

| w | 128 | 160 | 320 |
|---|---|---|---|
| A | 0.04 | 0.04 | 0.037 |
| B | 0.206 | 0.201 | 0.177 |
| C | 0.288 | 0.28 | 0.24 |

Table 2 Coefficients *A*, *B* and *C*

| KA | NA | KB | NB | KC | NC |
|---|---|---|---|---|---|
| -0.00002 | 0.043 | -0.00015 | 0.225 | -0.00025 | 0.32 |

Table 3 Values of linear coefficients *KA*, *NA*, *KB*, *NB*, *KC* and *NC* of coefficients *A*, *B* and *C*

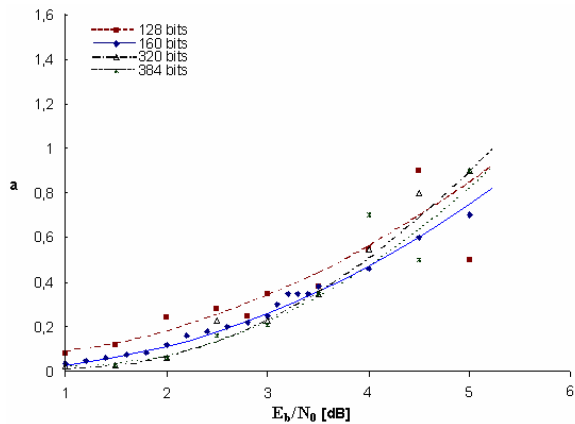

Fig. 15 Coefficient *a* in dependence on $E_b/N_0$ for different lengths of SID block

Using equations (7) and (9), the percentage of corrected bits (indicated as y in (7)) can be calculated as a function of positions of |L|-values (indicated as *x* in (7)), *S/N* and the length w of a SID block:

$$y = ke^{-ax} = \frac{e^{a(S/N,w)} - 1}{1 - e^{-a(S/N,w)w}} e^{-a(S/N,w)x} \quad (15)$$

with

$$a\left(\frac{S}{N}, w\right) = A(w) \cdot \left(\frac{S}{N}\right)^2 + B(w) \cdot \left(\frac{S}{N}\right) + C(w) =$$
$$(KA \cdot w + NA) \cdot \left(\frac{S}{N}\right)^2 + (KB \cdot w + NB) \cdot \left(\frac{S}{N}\right) + (KC \cdot w + NC) \quad (16)$$

If the number of |*L*|-values needed for correction are indicated by *x*, a normalized distribution function of percentage of correction by $x^{th}$ *L*-value *y* can be expressed, using equation (16), as:

$$y \approx ae^{-a*x}/(1-e^{-aw}) = \left[(KA \cdot w + NA)\cdot\left(\frac{S}{N}\right)^2 + (KB \cdot w + NB)\cdot\left(\frac{S}{N}\right) + (KC \cdot w + NC)\right] \cdot$$
$$\cdot \exp\left\{-\left[(KA \cdot w + NA)\left(\frac{S}{N}\right)^2 + (KB \cdot w + NB)\left(\frac{S}{N}\right) + (KC \cdot w + NC)\right] \cdot x\right\} \Big/ \Big(1 -$$
$$-\exp\left\{-\left[(KA \cdot w + NA)\left(\frac{S}{N}\right)^2 + (KB \cdot w + NB)\cdot\left(\frac{S}{N}\right) + (KC \cdot w + NC)\right] \cdot w\right\}\Big)$$

(17)

If the number $x_0$ of *L*-values needed for correction of $y(x_0)$ percentage of errors has to be, equation (7) is expressed as:

$$y(x_0) \approx \sum_{i=1}^{x_0} ke^{-a*x} = \sum_{i=1}^{x_0} \frac{e^a - 1}{1 - e^{-aw}} e^{-ax} = \frac{1 - e^{-ax_0}}{1 - e^{-aw}} \quad (18)$$

and $x_0$ is found as:

$$x_0 \approx \frac{1}{a}\ln\left(\frac{1}{1 - y(x_0)(1 - e^{-aw})}\right) \quad (19)$$

It is noted that equation (19) is based on an interpolation of $E_b/N_0$ = [1, 4] dB (i.e. *S/N* = [-2, 2] dB) and *w* = [128, 160, 320].

$y(x_0)$ can be between 0 and 100 [%], but it is 95 % in this paper.

The maximum position of minimum |*L*|-value can be found for different lengths of SID blocks and different $E_b/N_0$.

In order to check equation (19), $y(x_0)$ (needed *L*-values) for correction of 95% SID blocks of length of 384 bits (extrapolation) are compared to the results of tests (Fig. 16):

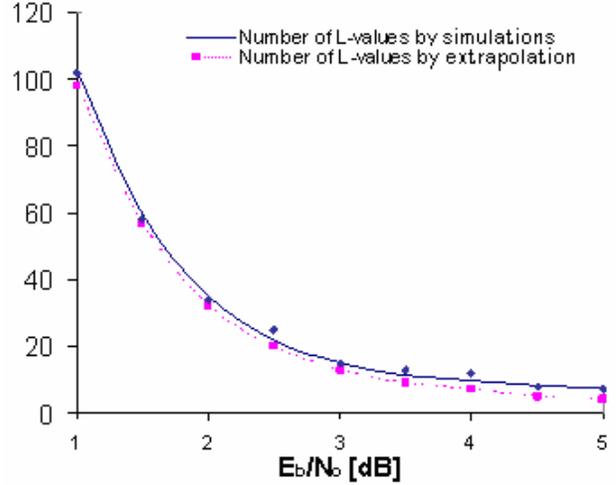

Fig. 16 Comparison of *L*-values needed for correction of 95% of SID blocks (*w* = 384) based on equation (19) and simulations (Fig. 10)

Fig. 16 shows that equation (19) can be trusted and used for further extrapolations.

Equation (19) is now applied for SID blocks of length *w* = 1024 in Fig. 17. The length of 1024 bits is chosen for comparison, because RSA signatures with a signature length of 1024 bits offer the same security level as digital signatures based on ECC with the signature length of 320 bits. It may appear attractive to use RSA signatures because the verification time is very short when a public exponent with a low Hamming weight is used.

Assuming $E_b/N_0$ = 3 dB, the number of *L*-values is 48 compared to 14 when using ECC digital signatures of 320 bit. Therefore, at one hand, it becomes clear, that the use of RSA digital signatures is not suitable for Soft Input Decryption. At the other hand, the application of Soft Input Decryption to longer blocks than 1024 bits is realistic if $E_b/N_0$ is not too low.

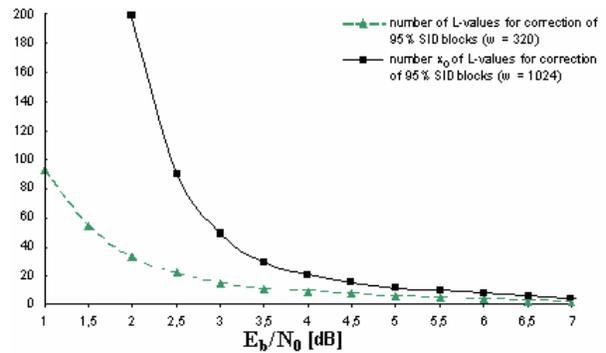

Fig. 17 Comparison of *L*-values needed for correction of 95% of SID blocks with *w* = 320 and *w* = 1024

Fig. 17 shows that 48 L-values rsp. $2^{48}$ tests are necessary in the case of RSA digital signatures are needed compared to 14 rsp. $2^{14}$ tests in case of ECC digital signatures.

## 7. Time Performance

Under practical aspects it is important to consider the consumed resources of Soft Input Decryption, specifically the time needed for the correction of SID blocks. The needed time depends on the cryptographic algorithm and on the number of *L*-values which are necessary to correct the errors. Therefore the success of Soft Input Decryption is highly dependent on the quality of the soft output of the channel decoder.

The time needed to perform $2^8$ verifications of digital signatures of 320 bits with appendix (based on ECDSA) without hash calculation which are not necessary in Scenario 2, MACs (CBC MAC – DES with key size of 56 bit) and H-MACs (SHA-1 with output size of 160 bit) are shown in Table 3 – 5. The numbers are based on use of a PC with Pentium 4 of 1.7 and 1.8 GHz [19].

The second row of the tables shows the number of verifications and consumed time, which is needed for testing and correcting bits with up to the 8 lowest |*L*|-values. In the case of MAC/H-MAC a new MAC/H-MAC is computed, if the changed bit is in the message part. If the changed bit is in the MAC/H-MAC part, only a comparison to the previously computed MAC/H-MAC is performed, which costs almost no time. It is assumed that the *L*-values used for correction are equally distributed over the message part and the cryptographic check value.

In the table with the results of digital signatures (Table 3) there is an additional third row which assumes an optimization of the digital signatures verifications (100 times faster verification) following the first verification. The arithmetic of subsequent verifications is much more efficient and faster, if only one bit of the input of the verification is changed compared to one of the preceding verifications because only a single point addition has to be executed for correction. Example: in the case of ECKDSA [13] only one single point addition is necessary for correction.

The most interesting result is that digital signatures are more suitable for Soft Input Decryption than MAC and H-MAC, if the assumption, that succeeding verifications need only 1 % of the first one, is true. Even if succeeding verifications need less than 25 % of the first one, Soft Input Decryption with digital signatures is faster than Soft Input Decryption with H-MACs or MACs.

|   | Number of Verifications | Time [*s*] 1,7 GHz | Time [*s*] 1,8 GHz |
|---|---|---|---|
| 1 | 1 | 0.0049 | 0.0044 |
| 2 | 256 | 1.2544 | 1.1264 |
| 3 | 1$^{st}$ + 255 similar | 0.0174 | 0.0156 |

Table 3 Computation times for verification of digital signatures

|   | Number of Verifications | Time [*s*] 1,7GHz | Time [*s*] 1,8 GHz |
|---|---|---|---|
| 1 | 1 | 0.002 | 0.00187 |
| 2 | 256 * *m* / (*m* + *n*) | 0.3072 | 0.2872 |

Table 4 Computation times for verification of MACs (*n* = 128 bits) of a message (*m* = 192 bits)

|   | Number of Verifications | Time [*s*] 1,7 GHz | Time [*s*] 1,8 GHz |
|---|---|---|---|
| 1 | 1 | 0.00424 | 0.004 |
| 2 | 256 * *m* / (*m* + *n*) | 0.543 | 0,512 |

Table 5 Computation times for verification of H-MACs (*n* = 160 bits) of a message (*m* = 160 bits)

## 8. Conclusion and Further Work

In this paper the principles of Soft Input Decryption are presented, as well as results of various simulations. The application of the Soft Input Decryption method is shown using different scenarios. The coding gain of cryptographic check error rates which is more than 2 dB in case of SID blocks of length of 320 bits show that Soft Input Decryption is a promising method which characteristics have to be examined for further use and improvement.

Different strategies for changing of bits are possible, depending on a used system and error distribution. In this paper a static strategy is used, but also other possible strategies as dynamic and BER strategy are mentioned. These and other suggestions for new strategies are for further study.

The performance of *L*-values is analyzed by comparison of results of tests and theoretical calculations. It is shown that the *L*-values are not reliable, because bits with lower |*L*|-

values are not necessarily wrong bits. For that reason much more *L*-values have to be tested, than it would be necessary if the wrong bits would really have the lowest |*L*|-values. Results show that *L*-values become less reliable with decreasing of $E_b/N_0$.

Existing results of performances of *L*-values (for SID blocks of length of 128, 160 and 320 bit) are used for generation of equations which approximately find the number of needed *L*-values for the correction of defined percentage of SID blocks. Results of approximation are confirmed by comparison to the known results of tests for SID blocks of length of 384 bits. By extrapolation of calculated equations on SID blocks of length 1024 bit, the number of needed *L*-values has been calculated, instead of performing of long Soft Input Decryption tests.

Time performance of Soft Input Decryption depends on used cryptographic mechanisms (digital signature, MAC or H-MAC). The surprising results is that digital signatures show very good performance results when the time needed for verification of "neighboured" signatures is improved.

The future work should focus on improvement of arithmetical efficiency of cryptography for Soft Input Decryption to achieve faster verification process in a scope of Soft input Decryption and to prove the assumption used in row 3 of Table 3.

Section 3.2 more sophisticated strategies of Soft Input Decryption instead of static strategy have been outlined. Further studies will elaborate these strategies and test the performance.

Investigation of Soft Input Decryption should be also applied to standardized turbo codes for 3G, using the fact that the remaining errors after turbo coding are very often grouped (flipping of corresponding groups of bits).

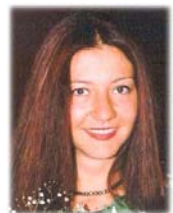

**Natasa Zivic, Dr.,** born 1975 in Belgrade, Serbia, graduated from the Faculty of Electrical Engineering (Electronics, Telecommunication and Automatics) of the Belgrade University in 1999. at the Telecommunication Department. After the Post diploma studies at the same Faculty (Telecommunications Division) she defended her Magister Thesis (Acoustics) in 2002.

From October 2004. she was scientific assistant at the University of Siegen in Germany at the Institute for Data Communications Systems as a DAAD and University of Siegen Scholarship holder. In 2007. she defended her Doctoral Thesis on the same University. The main course of her work in Siegen is Coding and Cryptography. From 2000. till 2004. she was working at the Public Enterprise of PTT "Serbia", Belgrade as the senior engineer. Currently she is employed as an Assistant Professor at the University of Siegen.